\documentclass[12pt]{JHEP3}
\usepackage{amsmath}
\usepackage{epsfig}

\psfull

\title{Non-global logarithms and jet algorithms in high-$p_T$ jet shapes}
\author{Andrea Banfi \\ Institute for Theoretical Physics, ETH Zurich, \\ 8093 Zurich, Switzerland \\
\email{banfi@itp.phys.ethz.ch}}
\author{Mrinal Dasgupta, Kamel Khelifa-Kerfa and Simone Marzani \\School of Physics \& Astronomy, University of Manchester,\\Oxford Road, Manchester, M13 9PL, U.K.
\email{mrinal.dasgupta@manchester.ac.uk\\ kamel.khelifa@hep.manchester.ac.uk \\simone.marzani@manchester.ac.uk }}

\preprint{MAN/HEP/2010/04}


\abstract{We consider jet-shape observables of the type proposed recently 
\cite{SCET1,SCET2} where the shapes of one or more high-$p_T$ jets, produced in a multi-jet event with definite jet multiplicity, may be measured leaving other jets in the event unmeasured. We point out the structure of the full next-to--leading logarithmic resummation specifically including resummation of non-global logarithms in the leading-$N_c$ limit and emphasising their properties. We also point out differences between jet algorithms in the context of soft gluon resummation for such observables.}

\begin{document} 




\section{Introduction}
It has long been known that probing the shape and structure of high-$p_T$ jets is potentially of great value in searches for new particles at collider experiments \cite{Seymour}. With the advent of the LHC and much activity in improving and  developing  jet algorithms~\cite{kt1}--\cite{SISCONE}, studies of this nature have received considerable impetus. In particular, much recent attention has been focused on using jet studies for the identification of boosted massive particles which decay to hadrons forming a collimated jet,  see for instance Refs~\cite{Butterworth:2002tt}--\cite{Chekanov:2010vc}.

In the same context a method has been recently suggested to study the shapes of
one or more jets produced in multi-jet events at fixed jet multiplicity
\cite{SCET1,SCET2}. The precise details of the observable suggested in those references involve defining a jet-shape energy-flow correlation similar to that
introduced in Ref.~\cite{SterBerg}. Specifically the proposal was to measure the shapes of one or more jets in an event leaving other jets unmeasured and introducing a cut on hadronic activity outside high-$p_T$ (hard) jets, to hold the hard-jet multiplicity fixed. This is in contrast to for instance hadronic event shapes \cite{Dasgupta:2003iq,BSZ01,BSZ04,BSZ10} which by construction are sensitive to the shape of the overall event rather than an isolated jet.

In the present paper we wish to use this observable as  a case study to make several points that we believe will be useful both within and outside the specific context. The main points that we wish to address concern the resummation of soft gluon effects that become important in describing the observable distribution for small values of the shape variable $\rho$ and the energy cut $E_0$.

In particular in this paper we shall address the  structure of non-global logarithms \cite{DassalNG1,DassalNG2} as well as compute them in the large-$N_c$ limit for jets defined in the anti-$k_t$ algorithm. We remind the reader that observables that are sensitive to radiation in a limited phase-space region such as the interior of a particular jet are non-global in the sense that they receive logarithmic contributions from correlated soft emission, which are highly non-trivial to treat to all-orders. Existing resummations of non-global logarithms have been confined to a few special cases \cite{DassalNG1,DassalNG2,Dassaldis,BMS,Rubin}
and to the large-$N_c$ limit. Given that the observable we study in the current paper is non-global, it is worth examining in detail the precise structure of non-global logarithms, which by definition start at $\mathcal{O}(\alpha_s^2)$ in the soft function and are of the same size as the logarithms resummed in  Refs.~\cite{SCET1,SCET2} and thus need to be considered as well in order to achieve NLL accuracy. In the current paper we find non-global logarithms arise both in  the ratio of the the energy cut-off $E_0$ and the shape variable $\rho$ as well as in $E_0/Q$ where $Q$ is the hard scale of the process, which is naturally of the order of the hard jet $p_T$\footnote{The potential presence of such logarithms was also mentioned in \cite{SCET1,SCET2}.}.  More to the point we argue that in the limit of narrow well-separated jets a simple picture emerges for non-global logarithms. The simplicity in the non-global structure is to do with the fact of QCD coherence. Narrow well separated jets do not affect each others evolution even in the non-global component which arises individually as an edge effect from the boundary of each jet, precisely as the non-global logs in the case of a hemisphere mass in $e^{+}e^{-}$ annihilation arise from the edge separating the observed and unobserved hemispheres~ \cite{DassalNG1}. Hence the resummation of non-global logarithms arising at each jet boundary can simply be  taken from the existing result for a hemisphere\footnote{This statement should be qualified as it is correct only for the case of the anti-$k_t$ jet algorithm~\cite{SalCac},  which is the one we recommend for study of such observables.} up to corrections that vanish as powers of the jet radius. The simple structure of non-global effects in turn provides us with an ansatz that can be used for any jet event of arbitrary jet multiplicity.

We also assess here the numerical contribution of the non-global logarithms and find that while limiting the value of $E_0$ is of some use in diminishing
their size the effect is still of order twenty percent as far as the peak height of shape distributions is concerned. 
In fact we find that changing the value of $E_0$ is not particularly useful as a means of reducing the non-global contribution. Specifically following the original proposal in Ref.~\cite{SterBerg} it was suggested in Ref.~\cite{SCET2} that one may take the value of $E_0/Q$ to be of the same order as the jet shape variable $\rho$, which we agree eliminates the non-global contribution from the measured jet. However in this case the contribution from the unmeasured jet becomes as significant as the contribution we are attempting to eliminate and hence the overall effect of this choice turns out to actually increase the overall non-global component. With the resummed results of the current article however one does not have to be too concerned about the precise choice of $E_0$ as the non-global terms should be accounted for, at least within the large-$N_c$ approximation and up to corrections vanishing as powers of the jet radius \footnote{These effects would amount to perhaps a ten percent change in the non-global term which we do not expect to be of significant phenomenological consequence.}.

Another point we wish to make is concerning the role of the jet algorithms.
The computation of non-global logs in the leading-$N_c$ limit can actually be carried out in any jet algorithm by means of the numerical codes developed for instance in \cite{DassalNG1,Rubin, SeyApp}. Indeed it was found \cite{SeyApp} that the use of certain sequential recombination algorithms (such as the $k_t$ or Cambridge-Aachen (C-A)) can significantly reduce the non-global logarithms due to the soft gluon clustering inherent in such algorithms. It was however later demonstrated \cite{BanDas05,BanDasDel} that one pays a price for this reduction in the non-global component in the form of a more complicated result for the independent emission terms. While independent emission is commonly associated with the exponentiation of the single-gluon result, this association is spoiled by the application of sequential recombination algorithms other than the anti-$k_t$ algorithm. As we shall show in this paper the result of soft gluon clustering in the $k_t$ and C-A algorithms modifies the independent emission (global term) which deviates from the naive exponentiation of a single gluon at a relevant single logarithmic accuracy. Moreover the effect of the clustering near the boundary of a collinear jet no longer produces logarithms suppressed in the jet radius $R$ as was the case for small central rapidity gaps discussed in Refs.~\cite{BanDas05,BanDasDel} but rather pure single-logarithms independent of $R$. These effects are absent for the anti-$k_t$ algorithm as already pointed out in Ref.~\cite{SalCac}, since that algorithm clusters soft gluons independently to the hard parton and hence produces circular jets in the soft limit, i.e. it can be regarded in this limit as a rigid cone.
Hence for the present moment and pending a resummation of the clustering logarithms along the lines of that carried out for gaps between jets \cite{BanDasDel} we confine our studies to the anti-$k_t$ algorithm. We do however provide an explicit fixed-order computation of the single-logarithmic corrections in the independent emission piece, that arise in other algorithms as we believe this point deserves some stress.

The paper is organised as follows. In Section~$2$ we define our observable, choosing the jet mass in dijet events as an example of a jet-shape observable, while imposing a cut $E_0$ in the inter-jet energy flow.
In Section~$3$ we perform the leading and next-to leading order calculation of such observable in the soft limit, which elucidates the structure of the logarithms arising from independent soft gluon emissions as well as non-global logarithms from correlated emissions. We use these results to construct an argument which culminates with the resummation of these logarithms in Section~$4$. We also present a study which assesses the numerical significance of the non-global logarithms as a function of the parameters $\rho$ and $E_0$. In Section~$5$ we highlight the fact that for algorithms other than the anti-$k_t$ exponentiation of the single gluon result is not sufficient to capture the next-to--leading logarithms \emph{even in the independent emission piece, let alone the non-global terms}. Finally we draw our conclusions in Section~$6$.

\section{High-$p_T$ jet shapes and inter-jet energy flow}
We wish to examine a situation where one studies the shapes of one or more high-$p_T$ jets in jet events with definite multiplicity. From the results we shall obtain below for such events one can draw conclusions also about the single inclusive jet mass distribution  for instance for the process $pp \to j+X$,
where one can demand the production of a jet $j$ setting a value for a particular jet-shape, while summing over everything else denoted by $X$.

For the points we wish to make in this paper we can for illustrative purposes and without loss of generality consider high-$p_T$ dijet events. In order to restrict the jet multiplicity we can place a cut $E_0$ whereby we veto the inter-jet activity such that the sum of transverse energies of emissions outside the two high-$p_T$ jets is less than this value. This was also the definition adopted in Refs.~\cite{SCET1,SCET2} where the parameter $\Lambda$ indicated a cut on additional jet activity along the above lines. 

Moreover, in this paper we are interested in physics at the boundary of the triggered hard jets and specifically in the non-global logarithms that arise at these boundaries. Hence we can for our discussion ignore the effects of initial state radiation which can simply be accommodated once the structure of the results is understood. Since it is this structure we wish to focus on, it proves advantageous to consider as an analogy the production of dijets in $e^{+}e^{-}$ annihilation which enables us to ignore the detail of initial state radiation.
Hence all our points can be made in full generality by considering two hard jets in $e^{+}e^{-}$ processes where one measures the shape of one of the jets leaving the other jet unmeasured as prescribed in Refs.~\cite{SCET1,SCET2}. Our results should also then be directly comparable to those obtained by other authors using soft-collinear effective theory \cite{SCET1,SCET2}. 

\subsection{Observable definition}
We shall pick the jet mass as a specific simple example of a jet-shape variable though one can consider also, for instance, the angularities first studied in \cite{SterBerg, BergMagnea}. 
The observable we study has the same logarithmic structure as the distribution in the angularity corresponding to $a=0$.
We study the shape cross-section
\begin{equation}
\label{eq:obs}
\Sigma\left(\rho,{E_0}\right) = \frac{1}{\sigma_0} \int \frac{d\sigma}{d \rho_1' d E_0'd^3 {\bf P_1} d^3 { \bf P_2}} d\rho' dE_0' \Theta(\rho-\rho_1') \Theta(E_0-E_0')\,,
\end{equation} where $\sigma_0$ is the Born cross-section and $\rho_1'$ denotes the normalised jet-mass of the jet with momentum $\bf{P_1}$. The above equation indicates that we are restricting the mass of the jet with three momentum $\bf{P_1}$ to be less than $\rho$ leaving the shape of the other jet with momentum $\bf{P_2}$ unmeasured. We have also restricted the inter-jet energy flow $E_0'$ to be less than $E_0$ as discussed and hence our observable definition above is in precise accordance with the definition in Refs.~\cite{SCET1,SCET2}. We shall in future leave the dependence on jet momenta $\bf{P_1},\bf{P_2}$ unspecified and to be understood. 

We wish to carry out a calculation for the above observable which includes a resummation of large logarithms in $R^2/\rho$ to next-to--leading logarithmic (equivalently single logarithmic)  accuracy  in the exponent. We include a description of non-global single logarithms in the leading-$N_c$ limit. Additionally we wish to resum the logarithmic dependence on $Q/E_0$,where $Q$ is the hard scale of the process, to single logarithmic accuracy again accounting for the non-global contributions. Our main aim is to study the effect of the non-global logarithms neglected for instance in previous calculations of jet shapes \cite{SCET1,SCET2,Sterman} on the cross-section Eq.~(\ref{eq:obs}). While resumming logarithms in $\rho$ and $E_0$ we shall neglect those logarithms that are suppressed by powers of the jet radius $R$  which shall enable us to treat non-global logarithms straightforwardly \footnote{More specifically we shall neglect corrections varying as $R^2/\Delta_{ij}$ where $\Delta_{ij}=1-\cos\theta_{ij}$ is a measure of the angular separation between the hard jets. This parameter emerges naturally in fixed-order computation of non-global logarithms for energy flow outside jets~\cite{Banfi:2003jj} and it was also treated as negligible in~\cite{SCET1,SCET2}.}. Hence our calculation addresses the range of study where $E_0/Q \gg \rho$ and is valid in the limit of relatively small jet radius $R$. We shall not however resum terms varying purely as $\alpha_s \ln R$ which for the values of $R$ we consider can safely be ignored from a phenomenological viewpoint. 
We thus aim to resum large logarithms in $\rho$ and $E_0/Q$ in what one may call the approximation of narrow well separated jets. According to our estimates this approximation and our consequent resummation should enable relatively accurate phenomenological studies of jet shapes.  

We shall begin by carrying out a calculation of the logarithmic structure that emerges at the one and two gluon levels, in the limit of soft gluon emission. 
These calculations will help us to identify the full logarithmic structure and point the way towards a resummed treatment. We start below with a leading order calculation in the soft limit. 

\section{Soft limit calculations}
We start by considering the effect of a single soft emission by a hard $q \bar{q}$ pair, produced in $e^+e^-$ annihilation. At this level all infrared and collinear (IRC) safe jet algorithms will yield the same result. We can write the parton momenta as 
\begin{align}
p_1 &= \frac{Q}{2}(1,0,0,1) \\
p_2 &= \frac{Q}{2}(1,0,0,-1) \\
k  & = \omega\left(1,\sin \theta \cos \phi, \sin \theta \sin \phi, \cos \theta \right)
\end{align}
where $p_1$ and $p_2$ are the hard partons and we have neglected recoil against the soft gluon emission $k$, which is irrelevant at the logarithmic accuracy we seek. Let us take the momentum $p_1$ to correspond to the measured jet direction. Hence if the gluon is combined with the parton $p_1$ one restricts the mass of the resulting jet to be below $\rho$ while if combined with $p_2$ the mass is unrestricted. Likewise one can consider the parton $p_2$ to be in the measured jet direction, which will give an identical result. 

Introducing the jet mass variable $\rho =M_j^2/E_j^2$, where one normalises the squared invariant mass $M_j^2$ to the jet energy $E_j^2$ one can write
\begin{equation}
\Theta\left(\rho -\frac{4 M_{j1}^2}{Q^2}\right)\Theta_{k \in j1}+\Theta\left(E_0-\omega \right) \Theta_{k \notin j1,j2}\,,
\end{equation}
where in our soft approximation the jet energy is set at $Q/2$. 
Note that there is no constrain on the gluon energy when it is combined with the jet $j2$.

In all the commonly used IRC safe jet algorithms the soft gluon $k$ will form a jet with a hard parton if it is within a specified distance $R$ of the hard parton. The distance is measured for hadron collider processes in the $(\eta,\phi)$ plane as $\Delta \eta^2 +\Delta \phi^2$ where $\Delta \eta$ is the separation in rapidity and $\Delta \phi$ is the separation in $\phi$ between the hard parton and the gluon $k$. In the limit of small angles, relevant for small $R$ values $R \ll 1$,  which we shall consider here, the distance measure reduces to $\theta_{pk}^2$ where $\theta_{pk}$ is the angle between the gluon $k$ and hard parton $p$.  Thus $k$ and $p_1$ form a jet if $\theta_{p_1k}^2<R^2$. Otherwise the gluon is outside the jet formed by $p_1$ which at this order remains massless. If the gluon does not also combine with hard parton $p_2$ to form a jet, one restricts its energy to be less than $E_0$ as required by the definition of the observable. Differences between the various algorithms shall emerge in the following section where we examine the emission of two soft gluons.

Thus we can write for the contribution of the real soft gluon $k$ with $\theta_{p_1k}^2 < R^2$ 
\begin{equation}
\Sigma^r = \frac{C_F \alpha_s}{\pi} \int 
\frac{d\omega}{\omega} \frac{d\theta^2}{\theta^2} \Theta \left(\rho -4 M_{j1}^2/Q^2 \right)\,.
\end{equation}
where we restricted the jet-mass to be less than the specified value $\rho Q^2/4$ and the superscript $r$ denotes the real emission piece.
In this same soft region virtual corrections are exactly
minus the real contributions, but unconstrained; therefore we can cancel the real emission result above entirely against the virtual piece and we are left with 
\begin{equation}
\label{eq;onegl}
\Sigma_{\mathrm{in}}  = -\frac{C_F \alpha_s}{\pi} \int_0^{Q/2}\frac{d\omega}{\omega} \int_0^{R^2}\frac{d \theta^2}{\theta^2} \Theta \left (\frac{2\omega \theta^2}{Q} -\rho \right)\,,
\end{equation}
where we constructed the jet mass $\rho= 4 \frac{\omega}{Q} (1-\cos \theta)\approx \frac{2\omega\theta^2}{Q}$, and where we used the small-angle approximation since $\theta^2 < R^2 \ll 1$.  The suffix ``$\mathrm{in}$'' denotes the contribution to $\Sigma$ from the region where the gluon is in the measured jet.
Performing the integral over angle with the specified constraint results in
\begin{equation}
\Sigma_{\mathrm{in}} = - \frac{C_F\alpha_s}{\pi} \int_{\rho Q/2R^2}^{Q/2} \frac{d\omega}{\omega} \ln \left (2\frac{\omega R^2}{Q \rho} \right) \\
=-\frac{C_F \alpha_s}{2\pi} \ln^2 \frac{R^2}{\rho} \Theta \left(R^2-\rho \right)\,.
\end{equation}

Next we consider the region where the soft emission flies outside either hard jets, with the corresponding contribution $\Sigma_{\mathrm{out}}$. Since here we are no longer confined to the small angle approximation we use $k_t$ and $\eta$ with respect to the jet axis as integration variables where $\eta$ is the gluon rapidity. In these terms one can represent the contribution of the gluon $k$ after real-virtual cancellation as 
\begin{equation}
\Sigma_{\rm out} = -\frac{2 C_F \alpha_s}{\pi} 
\int  \frac{dk_t}{k_t} \int_{-\ln 2/R}^{\ln 2/R}d\eta \, \Theta \left(k_t \cosh \eta - E_0\right)\,.
\end{equation}
where the limits on the rapidity integral reflect the out of jet region.
Performing the integrals we get to the required single-logarithmic accuracy
\begin{equation}
\Sigma_{\mathrm{out}}= -2 C_F \frac{\alpha_s}{\pi} \ln \left(Q/E_0\right) \left(2 \ln \frac{2}{R}\right)\,.
\end{equation}

The full soft result at leading order is $\Sigma_1 = \Sigma_{\mathrm{in}}+\Sigma_{\mathrm{out}}$. As is well known the jet-mass distribution receives double logarithmic corrections which in the present case are in the ratio $R^2/\rho$. 
Taking account of hard collinear emissions one would obtain also single logarithms in $R^2/\rho$, which we shall account for in our final results. 

The above calculation having set the scene we shall now move to considering two-gluon emission and the structure of the non-global logarithms that arise at this level.

\subsection{Two-gluon calculation and non-global logarithms}\label{non-global} 
Going beyond a single soft emission to the two gluon emission case the precise details of the jet algorithm start to become important. In what follows below we shall consider only the anti-$k_t$ algorithm since in the soft limit the algorithm functions essentially as a perfect cone algorithm \cite{SalCac}. 
In particular this implies that soft gluons are recombined with the hard partons independently of one another (one can neglect soft gluon clustering effects)
which considerably eases the path to a resummed prediction. The logarithmic structure for other jet algorithms is also interesting and we shall discuss it in a later section.

Here we carry out an explicit two-gluon calculation to obtain the structure of non-global logarithms for the observable at hand. 
 Referring to the non-global contribution to $\Sigma$ as $S$, we compute below $S_2$ the first non-trivial term of $S$.
Our results shall indicate a way forward towards a resummed result incorporating these effects.
As in  Refs.~\cite{DassalNG1,DassalNG2} we shall consider the emission of gluons $k_1$ and $k_2$ such that $\omega_1 \gg \omega_2$, i.e. strong energy ordering.
In this limit the squared matrix element can be split into an independent 
emission term $\propto C_F^2$ and a correlated emission term $\propto C_F C_A$
 The former is incorporated in the standard resummed results based on exponentiation of a single gluon, which we discuss later.

Let us concentrate on the $C_F C_A$ term missed by the single gluon exponentiation, and which generates the non-global logarithms we wish to study and resum. 
We now consider the following kinematics:
\begin{align}
p_1 &=\frac{Q}{2} \left(1,0,0,1\right) \\ \nonumber
p_2 &= \frac{Q}{2} \left(1,0,0,-1 \right) \\ \nonumber
k_1 &= \omega_1 \left(1,\sin\theta_1,0,\cos\theta_1\right)\\ \nonumber
k_2 &= \omega_2 \left(1,\sin\theta_2 \cos\phi, \sin \theta_2 \sin \phi, \cos \theta_2 \right) \\
\end{align}
with $\omega_1 \gg \omega_2$.

\begin{figure}
\begin{center}
\epsfig{file=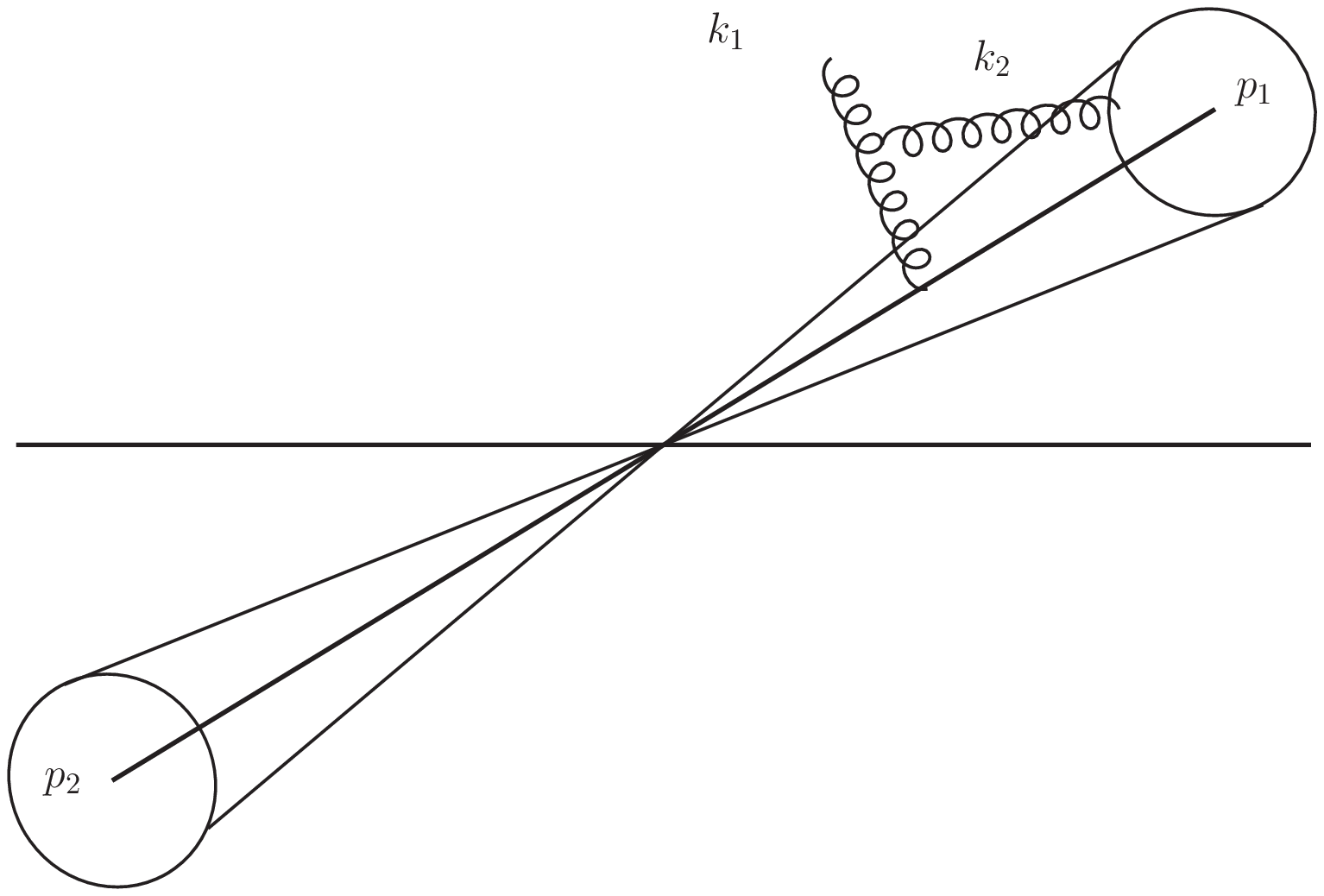, width = 0.45 \textwidth}
\hspace{0.5cm}\epsfig{file=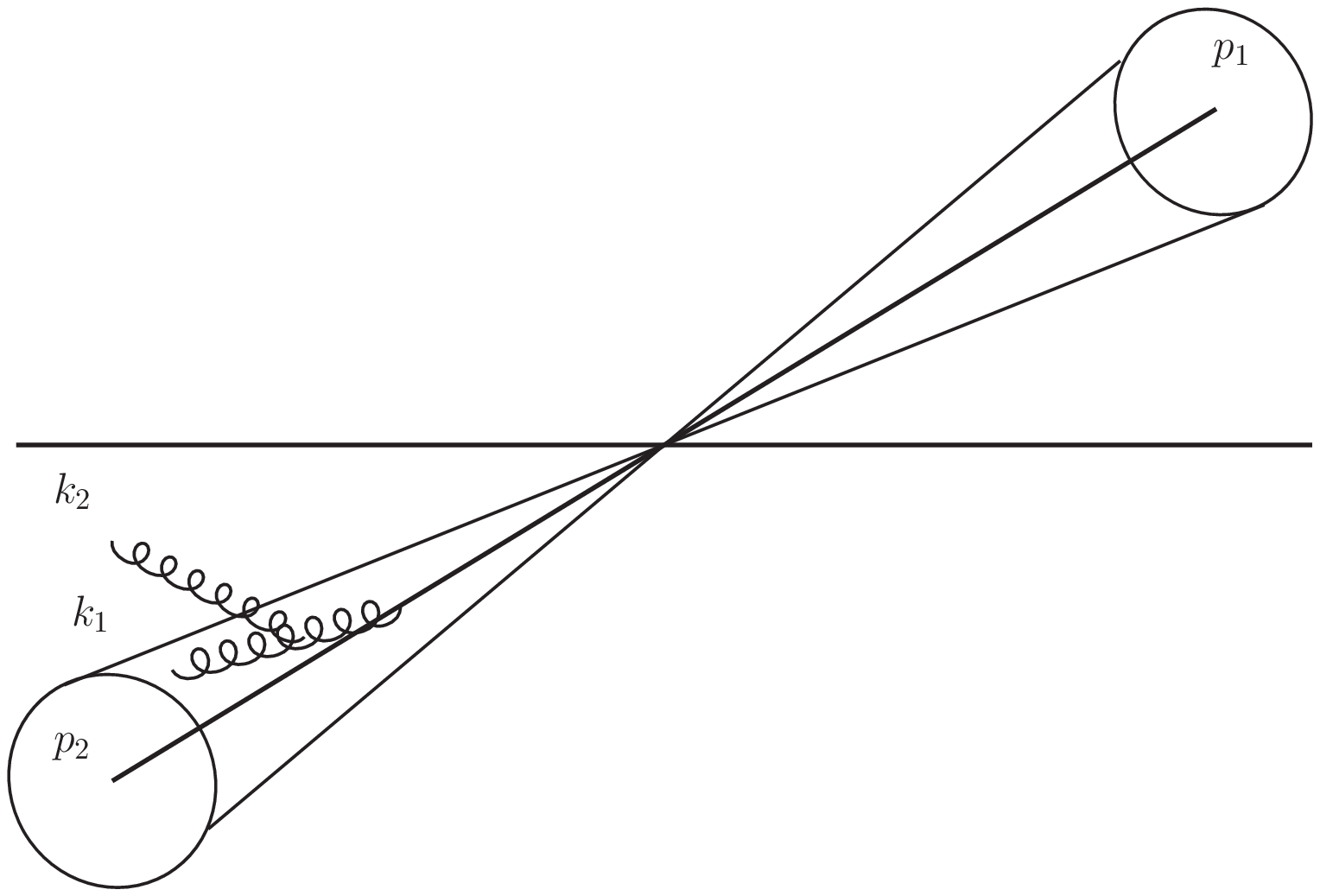, width = 0.45 \textwidth}
\caption{Diagrams representing the correlated emissions which give rise to the lowest-order non-global logarithms. On the left: the harder gluon $k_1$ lies outside both jets and the softest one $k_2$ is recombined with the measured jet and contributes to the jet-mass distribution. On the right: the harder gluon is inside the unmeasured jet and emits a softer gluon outside both jets, which contributes to the $E_0$-distribution. }
\label{fig:non-globalcontribution}
\end{center}
\end{figure}

Let us consider the situation where the harder gluon $k_1$ is not recombined with either jet but the softest emission $k_2$ is recombined with $p_1$. This situation corresponds to the diagram of the left in Fig.~\ref{fig:non-globalcontribution}.  In the small-angle limit, which applies for the case $R \ll 1$, the condition for $k_2$ to be recombined with $p_1$ is simply $\theta_2^2< R^2$ or equivalently $1-R^2/2 < \cos\theta_2 < 1$ while one has $-1+R^2/2 < \cos\theta_1 <1-R^2/2$ which ensures that $k_1$ is outside the jets. We integrate the squared matrix element for ordered soft emission \cite{DMO} over the azimuth of gluon $k_2$ to get the angular function~\cite{DassalNG1}
\begin{equation}
\Omega = \frac{2}{\left(\cos\theta_2-\cos\theta_1 \right) \left(1-\cos\theta_1\right) \left(1+\cos \theta_2 \right)}.
\end{equation}
Then defining the energy fractions $x_i = \frac{2 \omega_i}{Q}$, the required integral for the non-global logs reads 
\begin{multline}
S_2 = -4 C_F C_A \left ( \frac{\alpha_s}{2\pi} \right)^2 \int_0^1 \frac{dx_1}{x_1} \int_0^1 \frac{dx_2}{x_2} \Theta \left(\frac{2E_0}{Q}-x_1\right)
\Theta\left(x_1-x_2 \right)
\\ \int_{1-R^2/2}^{1} d\cos\theta_2 \int_{-1+R^2/2}^{1-R^2/2} d\cos\theta_1  \, \Omega \, \Theta \left(2x_2 (1-\cos\theta_2)-\rho \right),
\end{multline}
where we note the constraints on $k_1$ and $k_2$ imposed by the observable definition. Note that as in Ref.~\cite{DassalNG1} the constraint on $k_2$ emerges after including the term where $k_2$ is a virtual gluon such that the divergence of real emission is cancelled and the piece we retain above is the virtual leftover. Integrating over $x_1, x_2$ we obtain
\begin{multline}
S_2 = - 2 C_F C_A \left ( \frac{\alpha_s}{2\pi} \right)^2 \int_{1-R^2/2}^{1} d\cos\theta_2 \int_{-1+R^2/2}^{1-R^2/2} d\cos\theta_1  \, 
\\ \ln^2  \frac{\rho Q}{4E_0(1-\cos\theta_2)} \Theta\left(1-\frac{\rho Q}{4E_0(1-\cos\theta_2)}\right)\,\Omega\,.
\end{multline}

The angular integrations  over $\theta_1,\theta_2$ gives the number 
$\pi^2/6$ provided we neglect terms of order $R^2$ and those varying as $\rho Q/(2 E_0 R^2)$. 
We recall that as stated before we neglect logarithms suppressed by powers of $R$ and also that our resummation will be valid when $\rho/R^2 \ll E_0/Q$ and hence can ignore the corrections to $\pi^2/6$.

Thus in the small $R$ limit the result for the leading 
non-global piece is 
\begin{equation}
\label{eq:ngrho}
S_2 = -C_F C_A \left (\frac{\alpha_s}{2\pi}\right)^2 \frac{\pi^2}{3} \ln^2\frac{2 E_0 R^2}{\rho Q} \Theta \left(\frac{2 E_0 R^2}{Q}-\rho \right).
\end{equation}

We note that one also can receive a contribution to the non-global logs from the case where $k_1$ is part of the unmeasured jet but this configuration produces a coefficient that varies as $R^2$ and hence can be ignored, consistently with our approximation. 
Lastly carrying out the integration with the harder gluon $k_1$ inside the measured jet and the softest one $k_2$ outside does not give us large logarithms in the region we are interested in, hence Eq.~\eqref{eq:ngrho} is our final result for the first non-global piece affecting the $\rho$ distribution.

Next we consider the case that the harder gluon $k_1$ is in the unobserved jet and emits $k_2$ outside both jets, as depicted in Fig.~\ref{fig:non-globalcontribution}, on the right.  In this case repeating the calculation in the same way produces to our accuracy 
\begin{equation}
\label{eq:nglam}
S_2 = -C_FC_A \left (\frac{\alpha_s}{2\pi}\right)^2 \frac{\pi^2}{3} \ln^2\frac{Q}{2 E_0}.
\end{equation}

The above results are noteworthy in many respects. Note that the result Eq.~\eqref{eq:ngrho} corresponds to the result already obtained for the hemisphere jet mass in
 Ref.~\cite{DassalNG1} provided one replaces $1/\rho$ in that result by $\frac{2E_0}{Q}/\frac{\rho}{R^2}$. This is because the non-global evolution takes place from energies of order $Q \rho/R^2$ up to those of order $E_0$, whereas for the hemisphere mass the relevant energy for the harder gluon was of order $Q$. 
More interestingly the coefficient of $S_2$, $\pi^2/3$, is the same as was obtained there. The origin of this is the fact that the collinear singularity between $k_1$ and $k_2$ dominates the angular integral. As has been noted before \cite{DassalNG2} as one separates the gluons in rapidity  the contribution to the non-global term, which represents correlated gluon emission, 
falls exponentially as gluons widely separated in rapidity are emitted essentially independently. Thus in our present case, up to corrections suppressed by $R^2$ the results for the $\rho$ distribution arise from the edge of the measured jet independently of the evolution of the unobserved jet. Likewise there are non-global logarithms given by Eq.~(\ref{eq:nglam}) which affect purely the inter-jet energy flow $E_0$ distribution. These arise purely from the edge of the unmeasured jets and are independent of the evolution of the measured jet which is well separated in rapidity (similar results were obtained in the work of Refs.~\cite{Rubin,Banfi:2003jj}). 

Thus a simple picture arises for non-global logarithms where each jet evolves independently and the effects arise from the edges of the jet with logarithms involving the ratio of the shape variable $\rho/R^2$ to the energy flow variable $E_0/Q$ coming from measured jets and unmeasured jets independently contributing logarithms in $Q/E_0$. The coefficients of the non-global logarithms will be identical within our accuracy to those computed for the hemisphere mass (where the effect is again an edge effect coming from the hemisphere boundary) and hence the resummation of the non-global effects from each jet can simply be taken from the resummation carried out in Ref.~\cite{DassalNG1} simply modifying the evolution variable. 
This will be done in the next section. 

To conclude we wish to draw attention to the fact that we have determined, with a fixed order calculation,  the precise non-global structure which was not included in Refs.~\cite{SCET1,SCET2} and 
knowledge of the nature of these logarithms should pave the way for more accurate phenomenological studies. We remind the reader that our study above is valid only for the case of the anti-$k_t$ algorithm. Other jet algorithms will give different non-global pieces as discussed in Refs.~\cite{SeyApp,BanDas05,BanDasDel}. In fact even the resummation of independent emission terms will be different in other algorithms, a fact that is not widely appreciated and that we shall stress in a later section. 

In the following section we turn to resummed results and provide a simple ansatz which will be valid for arbitrarily complex processes involving jet production.

\section{Resummation}
Having observed the key feature of the non-global logarithms (independent contributions from each jet) that allow us to write a resummed result we shall now focus on the resummation in more detail.
The main point to note is that the non-global logarithms provide a factor 
that corrects straightforward single-gluon exponentiation~\cite{DassalNG1}:
\begin{equation} \label{full}
\Sigma\left(\frac{R^2}{\rho},\frac{Q}{E_0}\right)=\Sigma^{ind}\left(\frac{R^2}{\rho},\frac{Q}{E_0} \right)  S^{ng}\left(\frac{E_0R^2}{Q\rho},\frac{Q}{ E_0}\right) \,.
\end{equation}
Thus we shall first provide the result for the single-gluon exponentiation taking account of hard-collinear emission and the running coupling, which contains leading and next-to--leading logarithms in $R^2/\rho$ as well as leading logarithms in $Q/E_0$.

\subsection{Independent emission contribution}
The resummation of independent emission contributions based on a squared matrix element that has a factorized structure for multi-gluon emission is by now a standard procedure and we shall avoid listing these details (see for instance \cite{CAESAR} for a detailed study of these techniques). We shall provide here only details of the final result for independent emission valid for the anti-$k_t$ algorithm only. We stress once again that even the independent emission piece will differ at next-to--leading logarithmic accuracy from that reported below if using another jet algorithm.

The result for the independent emission contribution can be written in the usual form~\cite{CTTW}
\begin{equation}
\Sigma^{ind}\left(\frac{R^2}{\rho},\frac{Q}{E_0}\right) = \frac{\exp 
\left[-\mathcal{R}_\rho -\gamma_E \mathcal{R}'_\rho \right]}{\Gamma\left(1+\mathcal{R}'_\rho\right)} \exp \left[-\mathcal{R}_{E_0}\right].
\end{equation}
Here $\mathcal{R}_\rho$ and $\mathcal{R}_{E_0}$ are functions of $R^2/\rho$ and $Q/E_0$ respectively, representing the exponentiation of the one gluon result. They describe the resummation of large logarithms to next-to--leading logarithmic accuracy in $R^2/\rho$ and leading logarithmic accuracy in $Q/E_0$ except for the inclusion of non-global logarithms not described by independent emission of soft gluons.

With inclusion of running coupling effects and the effects of hard collinear emission the function $\mathcal{R}_{\rho}$ can be written as 
\begin{equation}
\mathcal{R}_{\rho} = \frac{C_F}{\pi} \int \frac{dk_t^2}{k_t^2}\alpha_s(k_t) {\mathcal{F}}(k_t^2)\,,
\end{equation}
where we defined
\begin{multline}
{\mathcal{F}}(k_t^2) = \ln \left(\frac{QRe^{-\frac{3}{2}}}{2 k_t}\right) \Theta \left(\frac{QR}{2}-k_t\right)\Theta \left(\frac{k_t^2}{Q^2}-\frac{{\rho}}{4}\right)\\+\ln\left(\frac{2Rk_t}{\rho Q} \right)\Theta \left(\frac{{\rho}}{4}-\frac{k_t^2}{Q^2} \right)\Theta \left(\frac{k_t^2}{Q^2}-\frac{{\rho}^2}{4R^2}\right),
\end{multline}
where the factor $e^{-3/2}$ in the argument of the logarithm in the first term above takes account of the hard collinear region $2\omega/Q \to 1$ \footnote{In order to obtain this one replaces as usual $\frac{d x}{x} \to dx \frac{1+(1-x)^2}{2x}$ where $x=2 \omega/Q$ in the integral over gluon energy, which is essentially introducing the full splitting function instead of its soft singular term.}. 

Carrying out the integral over $k_t$ one obtains the familiar result for $\mathcal{R}_{\rho}$ as follows
\begin{equation}
\mathcal{R}_{\rho} = -Lf_1(\lambda)- f_2(\lambda) \,,
\end{equation}
and
\begin{equation}
\mathcal{R}'_{\rho} = - \frac{\partial}{\partial L } \left( L f_1(\lambda) \right) \,.
\end{equation}

The functions $f_1$ and $f_2$ are listed below 
\begin{equation}
\label{eq:quark}
f_1(\lambda) = - \frac{C_F}{2 \pi \beta_0 \lambda} \left [ \left(1-2 \lambda \right ) 
\ln \left(1-2\lambda \right)-2 \left ( 1-\lambda \right ) \ln \left
  (1-\lambda \right ) \right ],
\end{equation}
and
\begin{multline} 
f_2(\lambda) = - \frac{C_F K}{4 \pi^2 \beta_0^2} \left [2 \ln \left 
(1-\lambda \right ) - \ln \left (1-2 \lambda \right )\right ] - 
\frac{3 C_F}{4 \pi \beta_0} \ln \left ( 1-\lambda \right ) 
\\  -\frac{C_F \beta_1}{2 \pi \beta_0^3} \left [ \ln \left (1-2\lambda \right )-2 \ln 
\left (1-\lambda \right ) + \frac{1}{2} \ln^2 \left (1- 2 \lambda \right ) 
- \ln^2 \left (1-\lambda \right ) \right ],
\end{multline}
 $\lambda = \beta_0 \alpha_s L, \; L = \ln \frac{R^2}{\rho}$ and $\alpha_s =\alpha_s\left(Q R /2\right)$ is the $\overline{MS}$ strong coupling.
In the above results the $\beta$ function coefficients 
$\beta_0$ and $\beta_1$ are defined as
\begin{equation}
\beta_0 = \frac{11 C_A - 2 n_f }{12 \pi}, \; \beta_1 = \frac{17 C_A^2 - 5 C_A n_f -3 C_F n_f}{24 \pi^2}\,,
\end{equation}
and the constant $K$ is given by~\cite{CMW}
\begin{equation}
K = C_A \left (\frac{67}{18}- \frac{\pi^2}{6} \right ) - \frac{5}{9} n_f\,.
\end{equation}

Likewise for the function $\mathcal{R}_{E_0}$ we have 
\begin{equation}
\mathcal{R}_{E_0} = -\frac{2 C_F}{\pi \beta_0} \ln \frac{2}{R} \ln (1-2 \lambda)\,, \end{equation}
where here $\lambda =  \beta_0 \alpha_s L, \; L = \ln \frac{Q}{2 E_0} $ and $\alpha_s =\alpha_s\left(Q/2\right)$. Note that the function $\mathcal{R}_{\rho}$ contains both a leading logarithmic term $Lf_1(\lambda)$ and a next-to--leading or single logarithmic term $f_2 (\lambda)$ while the leading logarithms in $\mathcal{R}_{E_0}$ are single logarithms and next-to--leading logarithms in this piece are beyond our control. The term $\Gamma\left(1+\mathcal{R}'_{\rho}\right)$ arises as a result of multiple emissions contributing to a given value of the jet-mass  and is purely single-logarithmic. The corresponding function for the $E_0$ resummation would be beyond our accuracy and hence is not included. 
We note the results presented here for $\mathcal{R}_{\rho}$ are identical to the ones for the $e^+e^-$ hemisphere jet-mass~\cite{CTTW}, with the replacement $\rho \to \rho/R^2$ and $\alpha_s(Q) \to \alpha_s(Q R/2)$. 

\subsection{Non-global component}
The non-global terms arise independently from the boundary of individual 
jets in the approximation of narrow well-separated jets. 
The result for an individual jet is the same as that for energy flow into a semi-infinite rapidity interval which was computed in the large-$N_c$ limit in 
Ref.~\cite{DassalNG1}.

In our two-jet example the contribution of non-global logarithms can be thus be written as 
\begin{equation} \label{Sng}
S^{ng}\left(E_0R^2/(Q\rho),Q/E_0\right) = S\left(t_{\mathrm{measured}}\right) S\left(t_{\mathrm{unmeasured}}\right)\,,
\end{equation}
where the function $S(t)$ was computed in Ref.~\cite{DassalNG1}. From that reference one notes that
\begin{equation}
S(t) = \exp \left (-C_F C_A \frac{\pi^2}{3} \left(\frac{1+(at)^2}{1+(bt)^c} \right)t^2\right)\,,
\end{equation}
where $a=0.85C_A, \, b=0.86 C_A, \, c=1.33$. 

The single logarithmic evolution variables for the measured and unmeasured jet contributions read
\begin{eqnarray}
t_{\mathrm{measured}} &=&\frac{1}{2\pi} \int_{\frac{\rho Q}{2E_0 R^2}}^{1} \frac{dx}{x} \alpha_s(x E_0)\,,\\
t_{\mathrm{unmeasured}}&=& \frac{1}{2\pi} \int_{\frac{2E_0}{Q}}^1 \frac{dx}{x} \alpha_s\left(xQ/2\right)\,,
\end{eqnarray} 
which represent the evolution of the softest gluon with a running coupling that depends on the gluon energy. For the measured jet the softest gluon evolves between scales of order $Q \rho/R^2$ and $E_0$ while for the unmeasured jet the evolution is from $E_0$ up to the jet energy $Q/2$.

Carrying out the integrals (one-loop running coupling is sufficient here) gives 
\begin{eqnarray}
t_{\mathrm{measured}} &=&-\frac{1}{4\pi \beta_0} \ln\left(1-\beta_0 \alpha_s\left(E_0\right) \ln \frac{2E_0 R^2}{Q \rho} \right)\,, \\
t_{\mathrm{unmeasured}}&=& -\frac{1}{4\pi\beta_0} \ln \left(1-\beta_0 \alpha_s\left(Q/2\right)\ln \frac{Q}{2E_0}\right).
\end{eqnarray}
In the following sub-section we shall illustrate the effects of the non-global logarithms on the shape-variable distributions for different values of $E_0$. 

\subsection{Numerical studies}
Let us examine the impact of non-global logarithms on the differential jet mass distribution, divided by the inclusive rate;  at our level of accuracy we have:
 \begin{equation}
 \frac{1}{\sigma} \frac{{\rm d}  \sigma}{{\rm d} \rho}= \frac{{\rm d}  \Sigma}{{\rm d} \rho}\,,
 \end{equation}
with $\Sigma$ given by Eq.~(\ref{full}).
\begin{figure}
\begin{center}
\epsfig{file=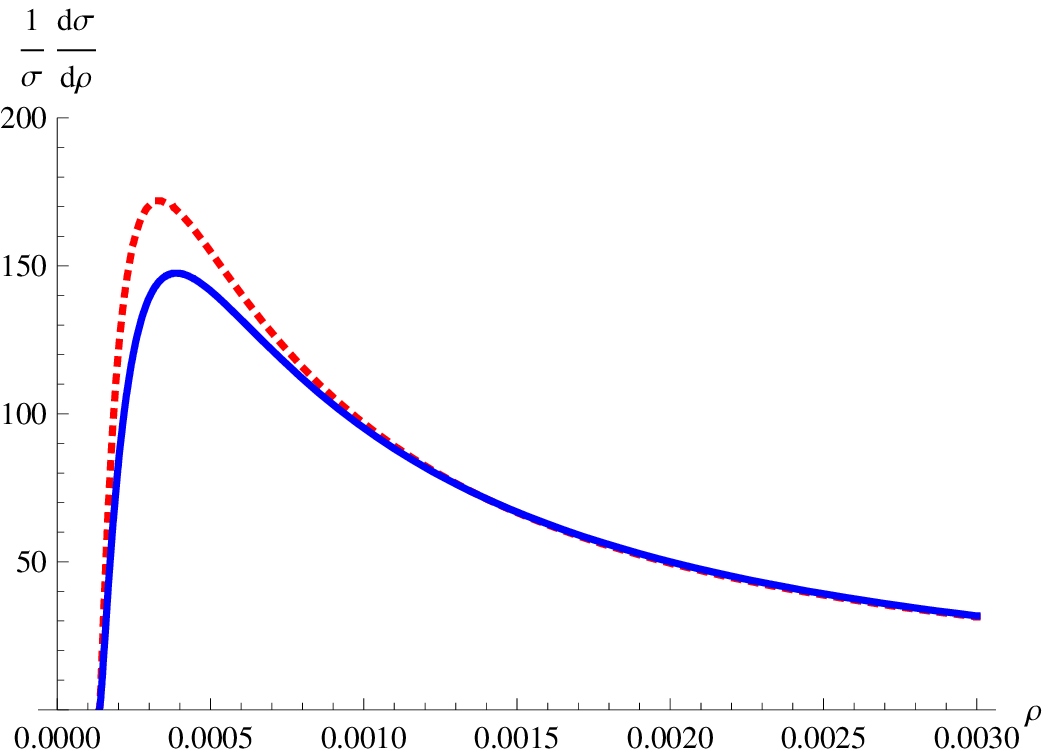, width = 0.48 \textwidth}
\epsfig{file=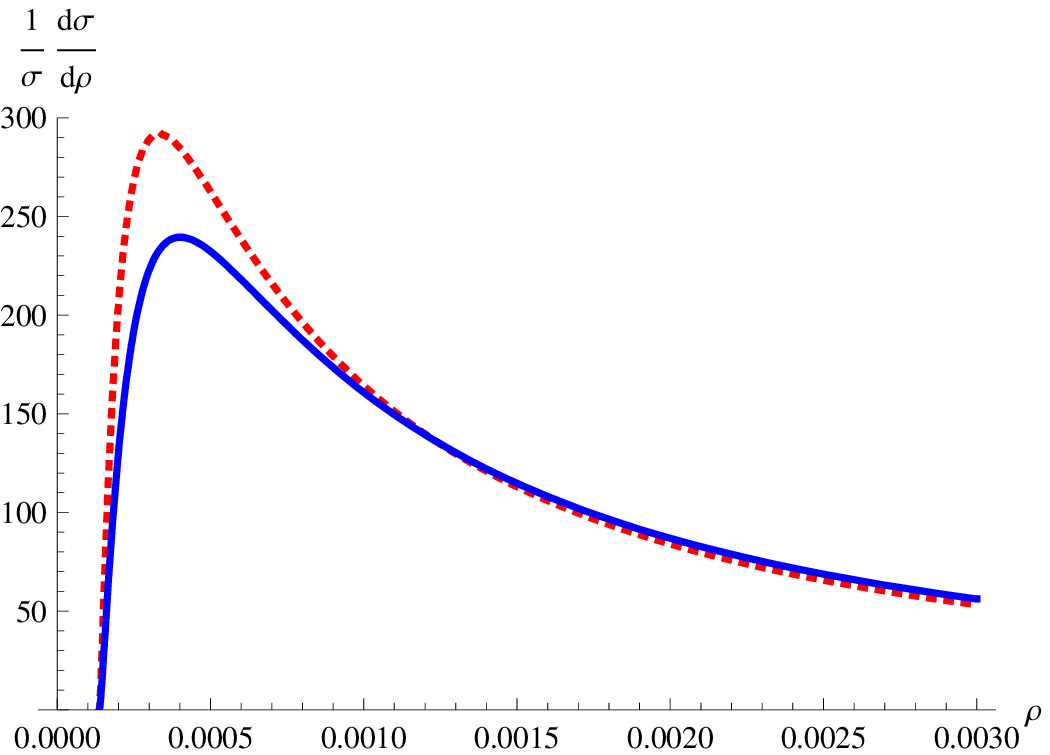, width = 0.48 \textwidth}
\caption{The jet mass distribution $\frac{1}{\sigma} \frac{{\rm d}  \sigma}{{\rm d} \rho}$ for 
$Q=500\, \mathrm{GeV}$, $R=0.4$ and $E_0 = 15 \,\mathrm{GeV}$ (left) and $60 \, \mathrm{GeV}$ (right). The curve in dotted red corresponds to neglecting non-global effects while that in solid blue takes them into account.}
\label{fig:non-global}
\end{center}
\end{figure}
From Figure.~\ref{fig:non-global} one can see that non-global logarithms do not change significantly the position of the peak of the distribution.  However, their inclusion leads to a reduction in the peak height of fifteen percent or so for $E_0 = 15\, \mathrm{GeV}$ and about twenty percent or so for $E_0 =60\,\mathrm{GeV}$. Increasing $E_0$ further one will observe that the effect of non-global logarithms on the peak height can be as significant as about $30 \%$. 
The plots above are for $Q=500 \,\mathrm{GeV}$ which may be translated into a jet $p_T$ of about $250 \,\mathrm{GeV}$ or so at a hadron collider. 

\begin{figure}
\begin{center}
\epsfig{file=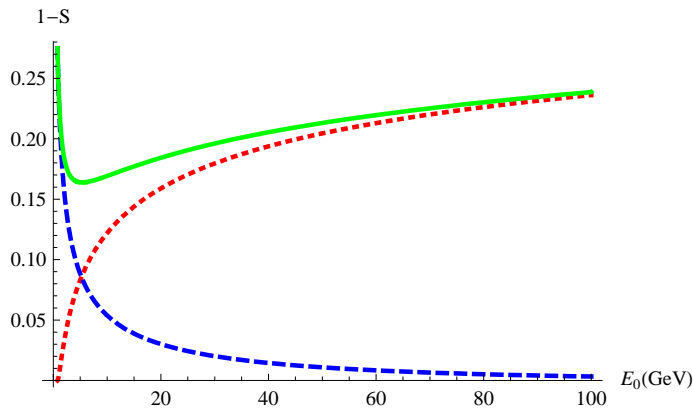, width = 0.6 \textwidth}
\caption{Non global contribution $1-S(t)$ from the measured jet (dotted red), the unmeasured jet (dashed blue) and overall (solid green) as a function of $E_0$ for $\rho = 5\times 10^{-4}, R=0.4$.} 
\label{fig:ngsplit}
\end{center}
\end{figure}

On the other hand it has been suggested \cite{SCET2} that one may eliminate non-global logs by choosing $E_0/Q$ of order $\rho$. In our case (small $R$) this prescription amounts to the choice $2E_0 R^2 = Q \rho$. While this rids us of non-global logarithms from the observed jet boundary the contribution from the unobserved jet boundary becomes increasingly important. This is reflected in Figure \ref{fig:ngsplit} which plots separately the factors $1-S(t_{\mathrm{measured}})$ (dotted red curve), $1-S(t_{\mathrm{unmeasured}})$ (dashed blue curve) and $1-S^{ng}$ (solid green curve), where $S^{ng}$ is defined in Eq.~(\ref{Sng}) as the product of the $S$ factors . The plots are presented as functions of $E_0$ for the illustrative value of $\rho = 5 \times 10^{-4}$. Other parameters are the same as for the previous plots. As one can readily observe increasing the value of $E_0$ leads to a growth of the non-global contribution from the measured jet while the contribution from the unmeasured jet is somewhat diminished. Lowering $E_0$ leads to the opposite effect and the unmeasured jet contributions become increasingly significant. It is noteworthy that changing the value of $E_0$ in the range indicated has no significant effect on the size of the non-global effect overall. Also worth noting however is that the choice $E_0 = \rho Q/(2R^2)$ (the lowest value of $E_0$ shown in the above mentioned plot) which eliminates the contribution from the measured jet (i.e. the red curve goes to zero) is not very helpful as the overall contribution stemming from the unmeasured jet entirely is more significant than for the higher values of $E_0$ discussed before. From this one realises that 
progressively decreasing the value of $E_0$ is not a way to eliminate the non-global contribution, for the observable at hand.

In the following section we shall show that for algorithms other than the anti-$k_t$ even the independent emission resummed result is not equivalent at next-to--leading logarithmic level to the exponentiation of the single-gluon result.

\section{Other jet algorithms}
Let us now consider the situation in other jet algorithms where the clustering or recombination of soft gluons amongst themselves may be an important effect. 
One such algorithm is the inclusive $k_t$ algorithm discussed for the case of central gaps between jets in Refs.~\cite{BanDas05, BanDasDel}. For such algorithms, starting from the two-gluon level, we need to revisit the independent emission calculations  and correct the naive exponentiation of a single gluon.  Note that in Refs.~\cite{BanDas05,BanDasDel} the single logarithms obtained as a result of clustering were proportional to powers of the jet radius which would make them beyond our control here. However, as we shall see, in the collinear region we are concerned with here, this power suppression does not emerge, making these logarithms relevant to our study. To illustrate the role of soft gluon clustering and recombination we focus on the on the single inclusive jet-mass distribution and we ignore the cut corresponding to $E_0$. Placing this cut does not affect the conclusions we draw here. 

\begin{figure}
\begin{center}
\epsfig{file=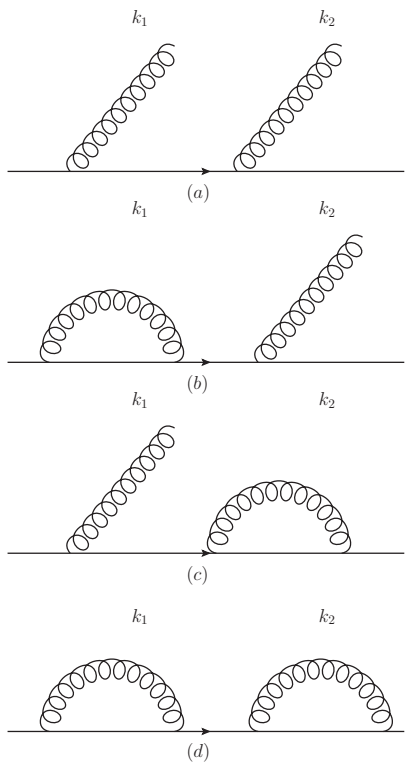, width = 0.35 \textwidth}
\end{center}
\caption{Diagrams contributing to independent two-gluon emission from a hard parton line.}
\label{fig:emission}
\end{figure}

To set the scene let us first carry out the independent emission calculation corresponding to two-gluon emission in the anti-$k_t$ algorithm which in the soft limit works like a perfect cone. At the two-gluon level we have four terms corresponding to 
the independent emission of soft gluons in the energy ordered regime $x_1 \gg x_2$. These contributions are depicted in figure \ref{fig:emission}.
The contribution to the squared matrix element for ordered two-gluon emission is the same for each of the diagrams in figure \ref{fig:emission}, up to a sign. The double real (labelled (a)) and double virtual contributions (labelled (d)) can be expressed as 
\begin{equation}
W(k_1,k_2) = 4 C_F^2 g^4 \frac{(p_1.p_2)^2}{(p_1.k_1)(p_1.k_2)(p_2.k_1)(p_2.k_2)}\,,
\end{equation}
which in terms of the energy fractions $x_1$ and $x_2$ introduced in section~\ref{non-global} becomes simply
\begin{equation}
W(k_1,k_2) = 256 g^4 \frac{C_F^2}{Q^4} \frac{1}{x_1^2 x_2^2} \frac{1}{\left(1-\cos^2\theta_1 \right)\left(1-\cos^2 \theta_2 \right)}\,.
\end{equation}
Since the calculation that follows below is intended for highly collimated jets, $R \ll 1$, we shall take the small angle limit of the above result, $\theta_1, \theta_2 \ll 1$. A similar result holds for the one-real one-virtual terms (b) and (c) in figure \ref{fig:emission} with a relative minus sign. We are now in a position to compute the jet mass distribution at the two gluon level for the independent emission $C_F^2$ term. 

We start by noting that the integration region for all graphs can be divided according to whether the real gluons $k_1$ and $k_2$ are inside or outside the triggered jet. We have four distinct regions: $k_1,k_2$ both outside the triggered jet, $k_1,k_2$ both inside the triggered jet or either of the gluons inside and the other outside the jet. The condition for a given gluon to end up inside or outside the triggered jet depends on the jet algorithm we choose to employ. 
In the anti-$k_t$ algorithm the condition is particularly simple when considering only soft emissions; such an emission $k$ is inside the jet if it is within an angle $R$ of the hard parton initiating the jet, else it is outside.

Given this fact let us consider how the various diagrams (a)--(d) in figure \ref{fig:emission} combine in the different regions mentioned above. 
Since we are computing the jet-mass distribution ${\rm d}\Sigma/{\rm d}\rho$ for a fixed jet-mass $\rho$, the pure virtual diagram (d) makes no contribution and hence we shall omit all reference to it in what follows.
In the region where both emissions are in the jet we shall treat the sum of graphs (a)--(c). Where the harder emission $k_1$ is in the jet and $k_2$ is out, graphs (a) and (c) cancel since the real $k_2$ does not contribute to the jet mass exactly like the virtual $k_2$. This leaves diagram (b) which gives zero since the in-jet gluon $k_1$ is virtual and hence does not generate a jet mass.
Hence the region with $k_1$ in and $k_2$ out gives no contribution. 

Now we consider $k_2$ in and $k_1$ out. The contributions with $k_2$ real (a) and (b) cancel as the graphs contribute in the same way to the jet mass. The diagram with $k_2$ virtual (c) cannot contribute to the jet mass as the real emission $k_1$ lies outside the jet. 

Hence we only need to treat the region with both gluons in and we shall show that this calculation correctly reproduces the result based on exponentiation of the single gluon result. The summed contribution of graphs (a) to (c) can be represented as 
\begin{equation}
\frac{{\rm d} \Sigma_{2}}{{\rm d}\rho} \sim \int d \Phi\, W \left[ \delta\left(\rho-x_1 \theta_1^2-x_2 \theta_2^2\right)-\delta\left(\rho-x_1 \theta_1^2\right)-\delta \left(\rho-x_2 \theta_2^2\right) \right]\,,
\end{equation}
where we wrote the contribution to the jet mass from an emission with energy fraction $x$ and angle $\theta$ with respect to the hard parton as $2x\left(1-\cos\theta \right) \approx  x \theta^2$. 

To compute the leading double-logarithmic contribution and show that it corresponds to the exponentiation of the order $\alpha_s$ double-logarithmic term one can write $\delta\left(\rho-x_1 \theta_1^2-x_2 \theta_2^2 \right)$ as $\frac{\partial}{\partial \rho} \Theta \left(\rho-x_1 \theta_1^2-x_2 \theta_2^2 \right)$ and make the leading-logarithmic approximation 
\begin{equation}
 \Theta \left(\rho-x_1 \theta_1^2-x_2 \theta_2^2 \right) \to \Theta \left(\rho-x_1 \theta_1^2 \right) \Theta\left(\rho-x_2 \theta_2^2 \right)\,,
 \end{equation}
 which allows us to make the replacement 
 \begin{equation}\delta\left(\rho-x_1 \theta_1^2-x_2 \theta_2^2\right) \to \delta\left(\rho-x_1 \theta_1^2\right)\Theta\left(\rho-x_2 \theta_2^2 \right)+1\leftrightarrow 2\,. \end{equation} 
Doing so and using the explicit forms of $W$ and the phase space $d \Phi$ in the small angle limit we get 
\begin{multline}
 \frac{{\rm d}\Sigma_{2}}{{\rm d}\rho} = - 4C_F^2 \left ( \frac{\alpha_s}{2\pi} \right)^2 
\int \frac{d\theta_1^2}{\theta_1^2}\frac{d\theta_2^2}{\theta_2^2} 
\frac{d\phi}{2\pi} \frac{dx_1}{x_1} \frac{dx_2}{x_2} \left[ \delta\left(\rho-x_1 \theta_1^2\right)\Theta\left(x_2 \theta_2^2 -\rho \right)+1 \leftrightarrow 2 \right] \\
\Theta \left(R^2-\theta_1^2 \right) \Theta\left(R^2-\theta_2^2\right) \Theta \left(x_1-x_2\right) \Theta(1-x_1).
\end{multline}
Carrying out the integrals we straightforwardly obtain
\begin{equation}
\frac{{\rm d}\Sigma_{2}}{{\rm d}\rho} = -2 \left(\frac{C_F \alpha_s}{2\pi} \right)^2 \frac{1}{\rho} \ln^3 \left (\frac{R^2}{\rho} \right),
\end{equation}
which is precisely the result obtained by expanding the exponentiated double-logarithmic one-gluon result to order $\alpha_s^2$ and differentiating with respect to $\rho$.
Thus the standard double-logarithmic result for the jet-mass distribution arises entirely from the region with both gluons in the jet. Contributions from soft emission arising from the other regions cancel in the sense that they produce no relevant logarithms. 

We shall now argue that for algorithms other than the anti-$k_t$ relevant single-logarithmic contributions shall appear from the regions which cancelled in the argument above, although of course the leading double-logarithms are still precisely the same as for the anti-$k_t$ case. An analysis of such miscancelling contributions is therefore necessary for a resummation aiming at next-to--leading logarithmic accuracy in the jet-mass. The logarithms we compute below correct the one-gluon exponentiated result for the jet-mass distribution at the single-logarithmic level starting from order $\alpha_s^2$. 

Let us consider the situation in, for instance, the $k_t$ algorithm. When both 
$k_1$ and $k_2$ are within an angle $R$ of the hard parton both soft gluons get combined into the hard jet and this region produces precisely the same result as the anti-$k_t$ algorithm, corresponding to exponentiation of the one-gluon 
result. Moreover, when both $k_1$ and $k_2$ are beyond an angle $R$ with respect to the hard parton there is no contribution from either to the jet-mass. However, when $k_1$ is beyond an angle $R$ and $k_2$ is inside an angle $R$ the situation changes from the anti-$k_t$ case. This is because in the $k_t$ algorithm when the two soft partons are separated by less than $R$ in angle they can be clustered together. The resulting soft jet has four-momentum $k_1+k_2$, when we use the four-momentum recombination scheme, and lies essentially along the harder gluon $k_1$. Thus when $k_1$ is beyond an angle $R$ it can pull $k_2$ out of the hard jet since the soft jet $k_1+k_2$ which replaces $k_2$ lies outside an angle $R$ of the hard parton. This results in a massless jet and hence such a configuration gives no contribution to the jet-mass distribution. In precisely the same angular region the virtual $k_1$, real $k_2$ diagram (b) (obviously unaffected by clustering) does however give a contribution whereas in the anti-$k_t$ case it had cancelled the double real contribution (a). The graph with $k_1$ real and $k_2$ virtual gives no contribution as before. Thus a new uncancelled contribution arises for the $k_t$ (and indeed the Cambridge--Aachen) algorithm from the region where the two real gluons $k_1$ and $k_2$ are clustered, which can be given by computing the $k_1$ virtual $k_2$ real graph in the same angular region.

We now carry out this calculation explicitly. We consider the angles $\theta_1^2$, $\theta_2^2$ and $\theta_{12}^2$ as the angles between $k_1$ and the hard parton, $k_2$ and the hard parton and $k_1$ and $k_2$ respectively. Applying the $k_t$ algorithm in inclusive mode means constructing the distances $\omega_1^2 \theta_1^2$, $\omega_2^2 \theta_2^2$ and $\omega_2^2 \theta_{12}^2$ along with the distances (from the ``beam'') $\omega_1^2 R^2, \omega_2^2 R^2$, where, for the $e^{+}e^{-}$ case we consider here, the energy $\omega$ plays the role of the $k_t$ with respect to the beam in a hadron collider event.
Now since $\theta_1^2 > R^2$, $\theta_2^2 <R^2$ the only quantities that can be a candidate for the smallest distance are $\omega_2^2 \theta_2^2$ and $\omega_2^2 \theta_{12}^2$. Thus the gluons are clustered and $k_2$ is pulled out of the jet if $\theta_{12}<\theta_2<R$. 
Otherwise $k_2$ is in the jet and cancels against virtual corrections. 

We can then write the contribution of graph (b) of Figure \ref{fig:emission} in the clustering region 
\begin{multline} 
\frac{ {\rm d}  } { {\rm d } \rho}\Sigma_2^{\mathrm{cluster} }=-4 C_F^2 \left(\frac{\alpha_s}{2\pi}\right)^2 \int \frac{d\theta_1^2}{\theta_1^2}\frac{d\theta_2^2}{\theta_2^2} \frac{d\phi}{2\pi} \frac{d x_1}{x_1} \frac{dx_2}{x_2} \delta \left(\rho -x_2 \theta_2^2 \right)\Theta(x_1-x_2) \\  \Theta\left(\theta_1^2-R^2\right) \Theta \left(\theta_2^2-\theta_{12}^2\right)\Theta \left(R^2-\theta_{2}^2\right).
\end{multline}
Using the fact that in the small-angle approximation relevant to our study 
\begin{equation} 
\theta_{12}^2 = \theta_1^2 +\theta_2^2-2 \theta_1 \theta_2 \cos \phi\,, 
\end{equation}
integrating over $x_1$ and $x_2$ and using $t= \frac{\theta_2^2}{\rho}$ 
one obtains
 \begin{multline} 
\frac{ {\rm d} } { {\rm d } \rho}\Sigma_2^{\mathrm{cluster} }= -4 C_F^2 \left( \frac{\alpha_s}{2\pi} \right)^2 \frac{1}{\rho} \int \frac{d\theta_1^2}{\theta_1^2} \frac{dt}{t} \frac{d \phi}{2 \pi} \ln t \\
 \Theta \left(t-1\right) \Theta \left(\theta_1^2-R^2\right) \Theta\left(4 \rho t \cos^2\phi-\theta_1^2\right)\Theta \left(R^2/\rho-t \right).
\end{multline}
Carrying out the integral over $\theta_1^2$ results in 
\begin{multline} 
\frac{ {\rm d}  } { {\rm d } \rho}\Sigma_2^{\mathrm{cluster} }= -4 C_F^2 \left( \frac{\alpha_s}{2\pi} \right)^2 \frac{1}{\rho} \int \frac{dt}{t} \frac{d \phi}{2 \pi} \ln \left(\frac{4 \rho t \cos^2 \phi}{R^2}\right) \ln t 
\\ \Theta \left(t-1\right) \Theta\left(4 \rho t \cos^2\phi-R^2\right)\Theta \left(R^2/\rho-t\right).
\end{multline}
Now we need to carry out the $t$ integral for which we note $t > \mathrm{max}\left(1, \frac{R^2}{4\rho\cos^2 \phi}\right)$. In the region of large logarithms 
which we resum one has however that $R^2 \gg \rho$ and hence $\frac{R^2}{4\rho\cos^2 \phi} >1$. This condition is reversed only when $\rho \sim R^2$ a region not enhanced by large logarithms and hence beyond our accuracy.

It is then straightforward to carry out the $t$ integral and doing so and 
extracting the leading singular behaviour in $\rho$ produces the result 
\begin{multline} 
\frac{ {\rm d}  } { {\rm d } \rho}\Sigma_2^{\mathrm{cluster} }
= -4 C_F^2 \left( \frac{\alpha_s}{2\pi} \right)^2 \frac{1}{\rho} \ln \frac{1}{\rho} \int \frac{d \phi}{\pi} \ln^2 \left(2 \cos \phi \right)\Theta \left (\cos \phi -\frac{1}{2} \right) \\ = -0.728 C_F^2 \left( \frac{\alpha_s}{2\pi} \right)^2 \frac{1}{\rho} \ln \frac{1}{\rho}.
\end{multline} 
This behaviour in the distribution translates into an next-to--leading logarithmic $\alpha_s^2 \ln^2 \frac{1}{\rho}$ behaviour in the integrated cross-section, which is relevant for resummations aiming at this accuracy. 
As we mentioned before the above finding of single logarithmic corrections generated by clustering has also been reported before for the case of gaps between jets studies \cite{BanDas05}. Note however that the logarithms found there had coefficients that depended on the jet radius as a power of the jet radius starting at the $R^3$ level. In the present case however the presence of collinear singularities near the boundary of a jet of small radius $R$ ensures that there is no power suppression in $R$ and hence the logarithms generated are formally comparable to those we aim to control here and indeed those resummed in Refs.~\cite{SCET1,SCET2}.
Likewise the clustering will also generate leading logarithms in the $E_0$ variable, which are again unsuppressed by any powers of $R$ and hence ought to be controlled. Lastly we point out that the logarithms generated by independent emission and subsequent $k_t$ clustering were actually resummed in Ref.~\cite{BanDasDel} and that possibility also exists here. 

\section{Conclusions}
We would like to conclude by emphasising the main points of our study. 
Given the current interest in the study of jet shapes and substructure for the purposes of discovering new physics at the LHC, it is worth examining the theoretical state of the art when it comes to looking at individual jet profiles in a multi-jet event. A step in this direction was taken for instance in Refs.~\cite{SCET1,SCET2}.
 In the present paper we have noted 
\begin{itemize}
\item Observables where one picks out for study one or more jets in multi-jet events are in principle non-global. The non-global logarithms will arise at next-to--leading or single-logarithmic accuracy in the jet-shape distributions.
If one studies jet events with a fixed multiplicity by imposing a cut $E_0$ on hadronic activity outside the high-$p_T$ jets, there are non-global logarithms 
involving the ratio of the shape variable $\rho$ and the energy flow $E_0$, as was first anticipated in \cite{SterBerg}. Moreover, there are also non-global logarithms in $E_0/Q$ where $Q$ is the hard scale of the process. These logarithms are leading as far as the distribution in $E_0$ for a fixed $\rho$ is concerned. 

\item In the limit of narrow jets $R \to 0$, one may naively expect the non-global contributions to the jet-shape distributions to vanish with $R$ due to the apparently limited phase-space available for soft emission inside the jet. Here  we have pointed out that the non-global logarithms do not vanish in the small cone approximation  as mentioned for instance in Ref.~\cite{deFlorian:2007fv}. One finds instead at small $R$ an effect that is independent of $R$ and arises from the edge of the jet. However, in the limit of narrow well-separated jets $R^2 \ll (1-\cos \theta_{ij})$, where $\theta_{ij}$ is the inter-jet separation, one can simplify the non-global contribution. In this limit, owing to QCD coherence and the nature of correlated multiple soft emissions, one can regard the non-global logarithms to arise independently from the boundary of each jet up to corrections that vanish as $R^2/\left(1-\cos \theta_{ij}\right)$. For a measured jet one picks up logarithms in $2E_0 R^2/(Q \rho)$ while for each unmeasured jet one has logarithms in $2E_0/Q$. The resummation of these logarithms yields a factor $S^j$ for each jet $j$, which is the factor computed, in the large-$N_c$ limit, for the hemisphere jet-mass in $e^{+}e^{-}$ annihilation in Ref.~\cite{DassalNG1}, again up to corrections vanishing as 
$R^2/\left(1-\cos \theta_{ij} \right)$. 

\item The overall size of non-global logarithms depends on the precise values one chooses for $E_0/Q$, $R$ and $\rho$. However, broadly speaking, we find the contribution not to vary significantly with $E_0$ and to yield corrections of order $15-20 \%$ in the peak region of the $\rho$ distribution. Choosing $E_0/Q$ of order $\rho/R^2$ eliminates the non-global contributions from the measured jet but steeply enhances the contributions from the unmeasured jet and it is not an optimal choice for reducing the overall non-global contribution to this observable.

\item We emphasise that the above observations are valid for the anti-$k_t$ algorithm in which our ansatz for resummation of jet shapes in an arbitrarily complex event is to correct the one-gluon exponentiation with a product of independent non-global factors from each jet. We further have emphasised that switching to algorithms other than the anti-$k_t$ gives relevant next-to--leading logarithms in the shape distribution as well as leading logarithms in the $E_0$ distribution, even within the independent emission approximation. Thus predictions for observables such as the one discussed in this paper, in those algorithms are prone to more uncertainty than our current study in the anti-$k_t$ algorithm at least until such logarithms are also resummed.
\end{itemize}

We would like to stress that the general observations in this paper are of applicability in a variety of other contexts. For instance, the issue of threshold resummation addresses limited energy flow outside hard jets, along the lines of the $E_0$ distribution here. The consequent non-global logarithms and the issue of the jet algorithm have not been addressed to any extent in the existing literature.  The same issues crop up in the case of resummation in the central  jet veto scale for the important study of Higgs production in association with two jets.

We hope that an awareness of the nature and size of the non-global contributions, the simplification that occurs in the small $R$ limit and our comments about the situation in other jet algorithms will help to generate more accurate phenomenological studies for these important observables at the LHC. In particular in future work we shall address in more detail the role of soft gluon effects and especially non-global logarithms on QCD predictions relevant to new physics searches at the LHC. As an existing example of such studies in the context of Higgs physics and the filtering analysis we can refer the reader to  Ref.~\cite{Rubin}. We shall aim to provide similar studies in the context of other shape variables in the near future. We also note that a study of resummed jet shapes and profiles would constitute an interesting test of QCD with early LHC data and we shall also generalise the present work with this aim in mind.

\section*{Acknowledgements}
We  thank the UK's STFC for financial support.

\end{document}